\newcommand{\tag}[1]{[{\it #1}]}
\newcommand{\ol}{\overline}
\newcommand{\kz}{Knizhnik-Zamolodchikov \relax}
\newcommand{\mmn}{\psi_{m_1, m_2, n}}

\documentstyle [12pt]{article}

\newlength{\extraspace}
\newlength{\extraspaces}
\newcommand{\beq}{\begin{equation}
\addtolength{\abovedisplayskip}{\extraspaces}
\addtolength{\belowdisplayskip}{\extraspaces}
\addtolength{\abovedisplayshortskip}{\extraspace}
\addtolength{\belowdisplayshortskip}{\extraspace}}
\newcommand{\eeq}{\end{equation}}

\newcommand{\beqa}{\begin{eqnarray}
\addtolength{\abovedisplayskip}{\extraspaces}
\addtolength{\belowdisplayskip}{\extraspaces}
\addtolength{\abovedisplayshortskip}{\extraspace}
\addtolength{\belowdisplayshortskip}{\extraspace}}
\newcommand{\eeqa}{\end{eqnarray}}

\newcommand {\dell}[1]{\frac{\partial}{\partial #1}}

\newcommand{\is}{\! & \! = \! & \!}
\newcommand{\n}{\nonumber \\[1.5mm]}
\newcommand{\half}{{1 \over{2}}}

\begin{document}

\begin{flushright}
{\sc NUS/HEP/}920503\\
May 1992
\end{flushright}
\vspace{.3cm}

\begin{center}
{\Large{\bf{Duality in Multi-layered Quantum Hall Systems
}}}\\[15mm]

{\sc Christopher Ting}\\[7mm]
{\it Defence Science Organization, 20 Science Park Drive, Singapore 0511
} \\[35mm]
{\bf Abstract}
\end{center}
\noindent
The braid group dynamics captures the fractional quantum Hall effect (FQHE)
as a manifestation of puncture phase. When the dynamics is generalized for
particles on a multi-sheeted surface, we obtain new tools which
determine the fractional charges, the
quantum statistics, and the filling factors of the multi-layered FQHE. A
many-quasi-hole wavefunction is proposed for the bilayered samples.
We also predict a $\nu = 5/7$ FQHE for triple-layered samples.
The viability of {\em 3-dimensional} FQHE and the application
of the concept of generalized duality to anyonic superconductivity are
discussed. \par
\vspace{1truecm}
\vfil

\newpage
\indent
Laughlin's theory \cite{Laughlin}
of fractional quantum Hall effect (FQHE) elegantly
describes the incompressibility of the 2-dimensional electronic system
under the influence of a strong, uniform magnetic field. The theory is
pivoted on an ans\"atz which is taken to be the ground state
$\psi_m$ of the many-body system displaying FQHE:
\beq
\psi_m
= \prod_{a < b} ( w_a - w_b )^m \exp (- \frac{1}{4 \ell^2} \sum_{a} | w_a |^2),
\label{L-ansatz}
\eeq
where $\ell$ denotes the magnetic length and $w_a$ are
the coordinates of $N$ electrons. Now, this ans\"atz is for
single-layered quantum Hall
system, with all the spins of the electrons polarized in the direction of the
magnetic field. Since electrons are fermions, in order to have
$\psi_m$ anti-symmetric with respect to the permutation of indices $a$,
$m$ must be an odd number. Recently, however,
two experimental groups \cite{Suen,Eisenstein}
successfully produced a $\nu = \half$ FQHE.
The samples they used were double-layered 2-dimensional electronic systems.
Labelling the two layers and treating
them as additional degrees of freedom, the $\nu = \half$ effect was
predicted by Yoshioka, MacDonald and Girven (YMG) \cite{YMG}. Basically, they
made a straightforward generalization of Laughlin's ans\"atz
(\ref{L-ansatz}) as follows:
\beqa
\mmn ( w_a^{\tag{1}}, \, w_a^{\tag{2}} )
= \! \! & & \! \! \prod_{a < b}^{N^{\tag{1}}}
(w_a^{\tag{1}} - w_b^{\tag{1}} )^{m_1}
\prod_{a < b}^{N^{\tag{2}}} (w_a^{\tag{2}} - w_a^{\tag{2}} )^{m_2}
\prod_{a, \, b}^{N^{\tag{1}}, N^{\tag{2}}}
( w_a^{\tag{1}} - w_b^{\tag{2}} )^n \n
\!\! & & \!\! \times \,
\exp  - \frac{1}{4 \ell^2} \left( \sum_a^{N^{\tag{1}}} |w_a^{\tag{1}}|^2
+ \sum_a^{N^{\tag{2}}} |w_a^{\tag{2}}|^2  \right) \, .
\label{YMG}
\eeqa
Presumably, the two labels $\tag{1}$ and $\tag{2}$ are some analogue of
spin, as $\mmn$ was first discussed by Halperin \cite{Halperin}
in the context of single-layered {\em un}polarized quantum Hall systems.
On the other hand, the novel thing about $\mmn$ in YMG's analysis is to
interpret $\prod_{ a, b} ( w_a^{\tag{1}} - w_b^{\tag{2}} )^n$
as inter-layer correlation. They concluded that $\mmn$ could well be a
good ans\"atz for the ground states of doubled-layered FQHE {\em if}
the distance $d$ between the layers is about 1.4 times of $\ell$ for
$\psi_{3, 3, 1}$ and $3 \ell$ for $\psi_{5, 5, 1}$.
To reveal the underlying physics of $\mmn$, it is important to find a
Hamiltonian for which $\mmn$ is an {\em exact} ground state and
to study the nature of the quasi-exictations of the double-layered
FQHE. What is the value of the fractional charge carried by each
quasi-hole? What is the quantum statistics of the quasi-holes?
Is it possible to have a theoretical tool that can generalize the analysis
of these issues even to {\em multi}-layered FQHE? These are the questions
we are going to answer.

Earlier, we have focussed on the {\em topological} underpinning of
the single-layered FQHE \cite{Ting-Lai-1,Ting-Lai-2}
from the braid group approach \cite{Lai-Ting}. It was understood that
Laughlin wavefunction $\psi_m$ satisfied the ground state equations of
a Hamiltonian suggested by the representation theory of the Artin braid group.
The braid group approach reveals that FQHE is a manifestation of the
puncture phase, wherein each electron sees its counterparts as punctures.
When the 2-dimensional systems enter the puncture phase, the
configuration space for each particle is no longer simply connected.
The non-trivial topology necessitates the consideration of homotopic
constraint that each possible path of a particle is imposed upon when
travelling from one location to the other. In this regard,
Y. S. Wu explicitly shows that the 1-dimensional path integral representation
of Artin braid group leads to a generalized quantum statistics interpolating
between bosons and fermions \cite{Wu}.
We introduced the concept of {\em charged}
winding ``number" and thereby expanded Wu's construction to include
the possibility of non-abelian anyons \cite{Lai-Ting,nonabelian}.
Beside being a non-abelian generalization of Wu's representation theory,
our formulation of homotopical constraint with the
charged winding number directly points out the relevance of \kz equations
\cite{KZ} to
the physics of anyons. Indeed, it can be readily seen that Laughlin's
wavefunction is an {\em exact} solution of
\beq
( D_a + \frac{B}{4} {\ol w}_a ) \psi_m = 0 \, ,
\label{GSE}
\eeq
where
\beq
D_a \equiv \dell{w_a} - m \sum_{b=1, b \neq a}^N \frac{1}{w_a - w_b}
\eeq
is the \kz operator for the conformal blocks of the current algebra.
We are working in a unit system where all the universal constants are
equal to 1 and $B$ is the magnetic field strength.
Using the same approach, we have also
captured $\mmn$ as an {\em exact} solution of
\beq
\left(
\dell{w_a^{\tag{1}}}
- m_1 \sum_{b \neq a}^{N^{\tag{1}}}
\frac{ 1 }{ w_a^{\tag{1}} - w_b^{\tag{1}}}
- n \sum_{b}^{N^{\tag{2}}}
\frac{ 1 }{ w_a^{\tag{1}} - w_b^{\tag{2}}}
+ \frac{B}{4} {\ol w}_a^{\tag{1}}
\right) \mmn = 0
\, .
\label{gse-double-a}
\eeq
\beq
\left(
\dell{w_a^{\tag{2}}}
- n \sum_{b}^{N^{\tag{1}}} \frac{ 1 }{ w_a^{\tag{2}} - w_b^{\tag{1}}}
- m_2 \sum_{b \neq a}^{N^{\tag{2}}} \frac{ 1 }{ w_a^{\tag{2}} - w_b^{\tag{2}}}
+ \frac{B}{4} {\ol w}_a^{\tag{2}}
\right) \mmn = 0
\, .
\label{gse-double-b}
\eeq
This may appear as a trivial generalization of (\ref{GSE}). The physics
behind, however, reaches much deeper than that \cite{mutual}.
It is shown that
(\ref{gse-double-a}, \ref{gse-double-b}) are the ground state equations
of particles living in the {\em double-sheeted} surface. The strength $n$
with which particles from the first sheet see those in the second sheet
as punctures can now differ from $m_1$, and vice versa.
As a consequence,
the possibility of {\em mutual} statistics between the particles of
different layers arises, as originally
discussed by Wilczek \cite{Wilczek-92} in the language of
adiabatic transport.
In order for $\mmn$ to be an anti-symmetric wavefunction, $m_1, m_2$
and $n$ are odd integers. This is because we are describing a system of
electrons, which are fermions. We realize that
the case with $n = m_1 = m_2$ is trivial.
This is simply the redundant way of saying that
the two layers are effectively indistinguishable and the two equations
(\ref{gse-double-a}, \ref{gse-double-b})
collapse back to (\ref{GSE}). Physically, it corresponds to the
situation where the distance $d$ between the layers is zero.
Therefore, for the Hamiltonian suggested by the
representation theory of braid group over the double-sheeted surface to
be a truely non-trivial generalization, the pathological case is
ruled out.
Next, one would expect $n$ to be no larger than $m_1$ and $m_2$. This is
because the inter-layer correlation should be weaker than the intra-layer
correlation. The strength $m_i$ with which particles of layer $i$ see
those of the same layer as punctures should be larger than $n$ with which
to see those of other layer. Let us summarize the physical arguments:
(A) $m_1$, $m_2$, $n$ are odd integers,
(B) $n = m_1 = m_2$ is forbidden, and
(C) $n \leq \min ( m_1, m_2) $.
A succinct way to express the relevant qunatum numbers of the double-layered
system is to write down a matrix of topological coupling strengths
after (\ref{gse-double-a}, \ref{gse-double-b}):
\beq
\left( \begin{array}{cc}
m_1 & n \\
n   & m_2
\end{array} \right)
\label{coupling matrix}
\eeq
Curiously, the physical inputs (A), (B) and (C)
imply a condition to be satisfied by (\ref{coupling matrix}):
$m_1 m_2 - n^2 > 0$.
Therefore, the coupling matrix for $\mmn$ is non-singular.

By looking at $\mmn$ alone, it is not clear how the unusual quantum statistics
may arise.
Afterall, since $m_1, m_2, n$ are odd integers, the quantum
statistics of the two layers of electrons characterized by $\mmn$
is still the usual one we know from the Pauli principle.
To appreciate the meaning of mutual statistics,
we have to consider the quasi-excitations. To warm up,
let us briefly sketch
how the quasi-hole wavefunctions of the single-layered systems
can be obtained from the braid group
dynamics. Suppose the charge of each quasi-hole is $c$, and we use
$u_a$, $a = 1, \cdots , N_h$ to denote their coordinates.
It turns out that the many-quasi-hole state $\psi_m^{qh}$
of $\psi_m$ satisfies the following equations:
\beqa
\left( \partial_{w_a}
+ \frac{B}{4} {\overline{w}_{a}}
- m \sum_{b=1, b \neq a}^N
\frac{1 \times 1}{w_a - w_b}
- m \sum_{b=1}^{N_h}
\frac{ 1 \times c }{w_a - u_b}
\right) \psi_m^{qh} \is 0 \, , \n
\left(
\partial_{u_a}
+ \frac{c B}{4} {\overline{u}_{a}}
- m \sum_{b=1}^{N}
\frac{ c \times 1}{u_a - w_b}
- m \sum_{b=1, b \neq a}^{N_h}
\frac{ c \times c }{u_a - u_b}
\right) \psi_m^{qh} \is 0 \, .
\label{qhole-gse}
\eeqa
One readily solves the equations and obtains
\beq
\psi_m^{qh}( u_1, \cdots, u_{N_h}; \, w_1, \cdots, w_N )
= \prod_{1 \leq a<b \leq N_h}
( u_a - u_b )^c
\exp (- \frac{c}{4 \ell^2} \sum_{a} | u_a |^2)
\prod_{a, b} ( u_a - w_b ) \, \psi_m
\label{qhole}
\eeq
It is well known that the quasi-holes are fractionally charged entities.
We shall see later how the value of $c$ is determined.
As can be seen from $(\ref{qhole})$, the statistics of the quasi-holes
is $\theta / \pi = c$.
Observe that there exists a kind of ``duality" in (\ref{qhole-gse}). Namely,
if we rewrite it in the following form
\beqa
\left( \partial_{w_a}
+ \frac{B}{4} {\overline{w}_{a}}
- m \sum_{b=1, b \neq a}^N
\frac{1 \times 1}{w_a - w_b}
- m c \sum_{b=1}^{N_h}
\frac{ 1 \times 1 }{w_a - u_b}
\right) \psi_m^{qh} \is 0 \, , \n
\left(
\partial_{u_a}
+ \frac{c B}{4} {\overline{u}_{a}}
- m c \sum_{b=1}^{N}
\frac{ 1 \times 1}{u_a - w_b}
- m c^2 \sum_{b=1, b \neq a}^{N_h}
\frac{ 1 \times 1 }{u_a - u_b}
\right) \psi_m^{qh} \is 0 \, .
\eeqa
we have a striking resemblance with the double-layered systems
(\ref{gse-double-a}, \ref{gse-double-b}).
In words, the quasi-holes of charge $c$ can be regarded as electrons
(of unit charge) living in the second layer! Viewed in this light,
the coupling matrix of such fictitious double-layered system is
\beqa
\begin{array}{ccccc}
\hspace{12mm} & e \hspace{3mm} & qh \nonumber
\end{array}
\eeqa
\vspace{-9mm}
\beq
\begin{array}{l}
e \n qh
\end{array}
\left( \begin{array}{cc}
m & mc  \\
mc & m c^2
\end{array} \right)
\label{coupling matrix e-qh}
\eeq
The crucial difference between two layers of real electrons and
one layer of electrons and quasi-holes is that while the coupling matrix of
the former is non-singular, that of the latter is singular.
That the coupling matrix for electron-quasi-hole system
(\ref{coupling matrix e-qh}) is singular is a manifestation of the
electric-magnetic duality. The charges $c$
of the quasi-holes are related to the
number of extra flux units being buffered by the Laughlin fluid when the
background magnetic field strength $B$ varies marginally from its
value for which the filling factor is exact. It turns out that $c$ is the
inverse of $m$, as the strength (magnetic charge) $m$ with which the
2-dimensional
system buffers a unit of extra flux is to create a quasi-hole of
(electric) charge $c$ so that $m c = 1$ flux unit.
We shall call (\ref{coupling matrix e-qh}) the associated coupling matrix
of $m$, and its being singular the auxiliary duality condition.
A remark is imperative at this juncture.
Strictly speaking, particles of different charges are distinguishable and the
conventional notion of statistics, which concerns a system of
particles being {\em in}distinguishable, does not apply {\em per se}.
But as the representation theory of braid group
does not require the charges to be identical,
one can still talk about two {\em distinguishable}
particles exchanging their locations. The so-called braid group ``statistics"
is topological in nature, for the configuration space for each particle is
riddled with punctures and hence non-simply connected.

Now let us generalize the duality concept encapsulated in $m c =1$ to the
double-layered systems:
\beq
\left( \begin{array}{cc}
m_1 & n \\
n   & m_2
\end{array} \right)
\left( \begin{array}{c}
c_1 \\
c_2
\end{array} \right)
= \left( \begin{array}{c}
1 \\
1
\end{array} \right) \, .
\label{matrix form}
\eeq
In other words, we have replaced $m$ by the coupling matrix
(\ref{coupling matrix}) and $c$ by the charge vector $(c_1, c_2)$.
In retrospect, the strength $m$ for the single-layered FQHE should be
considered as a $1 \times 1$ matrix and $c$ a $1 \times 1$ vector. The
duality condition determines the values of the charges $c_1, c_2$.
The meaning of (\ref{matrix form}) has been discussed in ref. \cite{mutual}
by counting the total flux that each electron sees. For electrons of layer
1, it is $\Phi^{\tag{1}} \equiv m_1 ( N^{\tag{1}} - 1 ) + n N^{\tag{2}} $,
while every electron of layer 2 sees a total of
$\Phi^{\tag{2}} \equiv m_2 ( N^{\tag{2}} - 1 ) + n N^{\tag{1}} $
units of flux. The crucial point is that at the large scale limit, the
two layers are in equilibrium and therefore
$\Phi^{\tag{1}} = \Phi^{\tag{2}}$, which is
\beq
(N^{\tag{1}} - 1 ) m_1 + N^{\tag{2}} n =
N^{\tag{1}} n + (N^{\tag{2}}-1) m_2 \, .
\label{equilibrium}
\eeq
Now, quasi-holes behave much like
electrons, except they are fractionally charged. Therefore, they also
engage themselves in the puncture phase. So,
in addition to $\Phi^{\tag{1}}$ that each electron of tag $\tag{1}$
sees, it also senses the flux carried by the quasi-holes:
$ m_1 c_1 N_1 + n c_2 N_2 = N_1 $.
Similarly, each electron of tag $\tag{2}$ senses an additional
$m_2 c_2 N_2 + n c_1 N_1 = N_2$
units of flux. Here, $N_1$ and $N_2$ denote the numbers of quasi-holes in
layer 1 and layer 2 respectively.
Since both layers are immersed in the same background
magnetic field, small variation in the field strength should result in
the same number of extra fluxes. In other words, $N_1 = N_2$.
Therefore, we get the following two equations:
$m_1 c_1 + n c_2  = 1$ and $m_2 c_2 + n c_1 = 1$.
These equations are none other than the generalized duality condition
(\ref{matrix form}). Upon solving the equations, we find
$c_1 = (m_2 - n) / (m_1 m_2 - n^2)$ and
$c_2 = (m_1 - n) / (m_1 m_2 - n^2)$.
Having determined the values of the charges of the quasi-holes, we
now calculate the filling factor of the double-layered FQHE.
The filling factor of the quantum Hall effect is given by
the total number of electrons divided by the total number of flux.
In the single-layered case, the total number of flux that an electron sees
is $(N - 1) m$. Hence, in the limit $N \rightarrow \infty$,
we have $\frac{N}{ ( N-1) m } \rightarrow \frac{1}{m}$.
Similarly, for the double-layered FQHE, using (\ref{equilibrium}) we get
$ (N^{\tag{1}} + N^{\tag{2}}) / (N^{\tag{1}} m_1 + N^{\tag{2}} n)$
$\rightarrow$ $(m_1 + m_2 - 2n) / (m_1 m_2 - n^2)$
in the limit $N^{\tag{1}} \rightarrow \infty$ and $N^{\tag{2}} \rightarrow
\infty$.
With the physical picture offered by the path integral representation of
the braid group,
we can proceed to write down the many-quasi-hole wavefunction of the
bilayered system:
\beqa
\psi & \equiv & \prod_{a < b} ( u_a^{\tag{1}} - u_b^{\tag{1}} )^{m_1 c_1^2}
\prod_{a, b} ( u_a^{\tag{1}} - u_b^{\tag{2}} )^{n c_1 c_2}
\prod_{a < b} ( u_a^{\tag{2}} - u_b^{\tag{2}} )^{m_2 c_2^2} \n
& & \times \,
\prod ( w_a^{\tag{1}} - u_b^{\tag{1}} )^{m_1 c_1}
\prod ( w_a^{\tag{2}} - u_b^{\tag{2}} )^{m_2 c_2}
\prod ( w_a^{\tag{1}} - u_b^{\tag{2}} )^{n c_2}
\prod ( w_a^{\tag{2}} - u_b^{\tag{1}} )^{n c_1} \n
& & \times \,
\exp ( - \frac{1}{4 \ell^2 c_1} \sum_a^{N_1} |u_a^{\tag{1}}|^2
- \frac{1}{4 \ell^2 c_2} \sum_a^{N_2} |u_a^{\tag{2}}|^2 )
\, \, \psi_{m_1, m_2, n}
\label{qh-bilayer}
\eeqa
This wavefunction is an exact solution of the bilayered version of
the ground state equations:
$\left( {\cal D}_a^{\tag{i}} + \frac{ c^{\tag{i}} B}{ 4 }
{\ol u}_a^{\tag{i}} \right) \, \psi  = 0 $ where
\beqa
{\cal D}_a^{\tag{1}} \!\! & \equiv & \!\!
\partial_{u^{\tag{1}}_a}
- m_1 \sum_{b}^{N^{\tag{1}}} \frac{c_1 \times 1}
{ u_a^{\tag{1}} - w_b^{\tag{1}} }
- n \sum_{b}^{N^{\tag{2}}} \frac{c_1 \times 1}
{ u_a^{\tag{1}} - w_b^{\tag{2}} }
- m_1 \sum_{b \neq a}^{N_1} \frac{ c_1 \times c_1 }
{ u_a^{\tag{1}} - u_b^{\tag{1}} }
-n \sum_{b}^{N_2} \frac{c_1 \times c_2}
{ u_a^{\tag{1}} - u_b^{\tag{2}} }
\n
\is \partial_{u^{\tag{1}}_a}
- m_1 c_1 \sum_{b}^{N^{\tag{1}}} \frac{ 1 \times 1}
{ u_a^{\tag{1}} - w_b^{\tag{1}} }
- n c_1 \sum_{b}^{N^{\tag{2}}} \frac{ 1 \times 1}
{ u_a^{\tag{1}} - w_b^{\tag{2}} }
\n
\! \! & & \! \! - \, \,
m_1 c_1^2 \sum_{b \neq a}^{N_1} \frac{ 1 \times 1}
{ u_a^{\tag{1}} - u_b^{\tag{1}} }
- n c_1 c_2 \sum_{b}^{N_2} \frac{1 \times 1}
{ u_a^{\tag{1}} - u_b^{\tag{2}} }
\, ,
\eeqa
and a similar expression for ${\cal D}_a^{\tag{2}}$ with
$ 1 \leftrightarrow 2$. The statistical parameters of the
quasi-holes are easily readable from (\ref{qh-bilayer}) and they are
\beq
\begin{array}{ccc}
\theta^{\tag{11}} / \pi = m_1 c_1^2 \, , \hspace{1cm} &
\theta^{\tag{12}} / \pi = n c_1 c_2 \, , \hspace{1cm} &
\theta^{\tag{11}} / \pi = m_2 c_2^2 \, .
\end{array}
\eeq
These values were first obtained by Wilczek \cite{Wilczek-92}
using the adiabatic transport argument.
We have checked that these values coincide with those calculated
with the generalized Chern-Simons approach \cite{genCS}.
$\theta^{\tag{12}} / \pi$ is the {\em mutual statistics}
of the quasi-holes residing in different layers.
The associated coupling matrix of the double-layered
systems is
\beqa
\begin{array}{ccccc}
\hspace{17mm} & e^{\tag{1}} \hspace{3mm} & e^{\tag{2}} \hspace{3mm}
& qh^{\tag{1}} \hspace{2mm} & qh^{\tag{2}} \nonumber
\end{array}
\eeqa
\vspace{-7mm}
\beq
\begin{array}{l}
e^{\tag{1}} \n e^{\tag{2}} \n qh^{\tag{1}} \n qh^{\tag{2}}
\end{array}
\left(
\begin{array}{cccc}
m_1     & n       & m_1 c_1   & n c_2   \n
n       & m_2     & n c_1     & m_2 c_2 \n
m_1 c_1 & n c_1   & m_1 c_1^2 & n c_1 c_2 \n
n c_2   & m_2 c_2 & n c_1 c_2 & m_2 c_2^2
\end{array}
\right)
\eeq
One can check that this $4 \times 4$ matrix is singular, as befits
the auxiliary duality condition for double-layered FQHE.

The generalization of the discussion to multi-sheeted FQHE is
immediate \cite{mutual}.
Suppose the number of layers is $M$. Then, the braid group dynamics
gives rise to a $M \times M$ coupling matrix which is non-singular,
and its associated $ 2 M \times 2 M$ matrix which is singular.
Using these matrices and the duality, we can characterize the
FQHE on multi-layered samples. In particular, one should get a
$\nu = 5 / 7$ for the triple-layered electronic systems with
coupling matrix
\beq
\left( \begin{array}{ccc}
3 & 1 & 0  \\
1 & 3 & 1  \\
0 & 1 & 3
\end{array} \right) \, .
\label{3 layers}
\eeq
Furthermore, it seems that one can even consider the following
coupling matrix for {\em 3-dimensional} FQHE:
\beq
\left( \begin{array}{cccccc}
m_1    &  n     &  0     & \cdots  &  \cdots   &  0      \\
n      &  m_2   &  n     & \cdots  &  \cdots   &  \vdots \\
0      &  n     & m_3    & \ddots  &           &  \vdots \\
\vdots & \vdots & \ddots & \ddots  &  \ddots   &  \vdots \\
\vdots & \vdots &        & \ddots  &  m_{M-1}  &  n      \\
0      &  0     & \cdots & \cdots  &  n        &  m_M
\end{array} \right)
\label{3-d}
\eeq
The symmetric tridiagonal coupling matrix (\ref{3-d})
is an interesting case because it corresponds to the
nearest-neighour inter-layer correlation which is more likely to
be realized experimentally. Now, when the number of layers $M$
is large, the electronic system is literally 3-dimensional!
By juxtaposing narrow
quantum wells alternately with barriers of appropriate width,
one may create 3-dimensional FQHE in the laboratory.
The multi-layered FQHE presented here through the looking glass of
coupling matrix may be relevant to the theory of anyonic superconductivity.
So far, most of the effective theories with a Chern-Simons piece in
the Lagrangian seem to ignore the
inter-plane correlation. It may well turn out that
the correlation between the copper oxide planes conspires to
circumvent the P and T violation \cite{Wilczek-92}. This is an open question.

In summary, we want to highlight the main theme of this work.
We see that the coupling matrix and the associated coupling matrix
contain all the relevant data that one needs to characterize the
FQHE in samples with arbitrary number of layers.
With the help of duality and the auxiliary duality condition,
we can determine the (fractional) charges of the quasi-hole excitations,
the statistics and the filling factor of the Hall media.
This set of tools comes from the representation theory of the
braid group over a multi-sheeted surface \cite{mutual}.

I would like to thank C. H. Lai for encouragement and discussions,
and C. L. Ho for a useful conversation on ``duality".

{\bf Note added}:
After this paper was submitted for publication, I learned
that Ezawa and Iwazaki had also independently considered
the concept of mutual statistics
(which they called relative statistics) in \cite{EI}.

\end{document}